# Electronic states, pseudo-spin, and transport in the zinc-blende quantum wells and wires with vanishing band gap

*J. B. Khurgin,[1]\* and I. Vurgaftman[2]*

*[1]Johns Hopkins University, Baltimore, MD 21218, USA*

*[2]Code 5613, Naval Research Laboratory, Washington DC 20375, USA*

## Abstract

We consider theoretically the electronic structure of quasi-two and quasi-one-dimensional heterostructures comprised of III-V and II-VI semiconductors such as InAs/GaInSb and HgCdTe. We show that not only a Dirac-like dispersion exists in these materials when the energy gap approaches zero, but that the states with opposite momentum are orthogonal (i.e. can be described by a pseudo-spin), which suppresses backscattering and thereby enhances the electron mobility, by analogy with the case of graphene. However, unlike in graphene, a quasi-one-dimensional quantum wire with zero gap can be realized, which should eliminate most of the scattering processes and lead to long coherence lengths required for both conventional and ballistic electronic devices.

\*Email: jakek@jhu.edu

Graphene, a single-layer honeycomb lattice of carbon, once capable of being produced in adequate quantities [1], has become a focus of interest in many fields, owing to its presumably unique properties [2] engendered by its gapless nature and the resulting Dirac-like linear dispersion. From the practical point of view, at least when it comes to electronic devices, the most interesting property of graphene is the reduced rate of scattering [3], which arises primarily from the nature of the Dirac states disallowing scattering into the opposite-momentum state. Backscattering is prohibited because the states with the opposite quasi-momenta, **k** and **–k**, have the orthogonal periodic Bloch (or "atomic") wavefunctions $\Psi_{\pm k} = \left( \Psi_A \pm \Psi_B \right) / \sqrt{2}$ , where $\Psi_{A(B)}$ is the P-type wavefunction of the C atom occupying sublattice A(B) of the graphene lattice with a two-atom basis. It is common to treat the states with the bonding $\left| A + B \right\rangle$ and anti-bonding $\left| A - B \right\rangle$ orbitals as two eigenstates of a pseudo-spin operator [4] that can be described just like an "ordinary" spin using the Pauli matrices. The introduction of pseudo-spin allows one to assign chirality to graphene, explain such fascinating phenomena as the Klein paradox [5], and has even led to speculation about pseudo-spin-based computing and information storage [6].

Since the Brillouin zone in graphene is two-dimensional (2D), the suppression of backscattering does not extend to scattering by smaller angles. As shown below, the mean momentum relaxation rate is only reduced by a factor of four corresponding to a four-fold increase in mobility. At the same time, in a 1D structure, the suppression of backscattering should dramatically improve the mobility and the mean free path. The original idea of suppressed backscattering in 1D quantum wires is due to Sakaki [7], who pointed out that backscattering becomes suppressed because it requires perturbation on the scale of $\approx \pi/k_F$, where $k_F$ is the Fermi wave vector. Interface roughness and ionized impurities with the characteristic scale larger than that cannot backscatter the electrons, and scattering by phonons with wavevector $q \approx 2k_F$ limits the mobility of these quantum wires. In single-wall metallic carbon nanotubes [8], graphene-like zero-gap dispersion does occur, and states **k** and –**k** have orthogonal periodic Bloch functions [9]; however, separating and fabricating devices from metallic nanotubes remains challenging.

Given these considerations, it is natural to consider whether the properties of graphene and nanotubes can be replicated in more conventional semiconductor systems. There has been a substantial amount of theoretical and experimental work on alternative zero-gap materials such as BiTe [10], BiSb [11], lead/tin salts, as well as many others, including the 2D structures of narrow-gap III-V (InAs/GaInSb) [12] and II-VI (HgTe/CdTe) [13-15] semiconductors. It was shown that quantum wells made of these materials can produce topological insulators in which such phenomena as Quantum Spin Hall Effect can be observed. In this article, we point out that if one can engineer 2D or 1D semiconductor structures with the vanishing (or very narrow) energy gap, the states with opposite **k** will exhibit orthogonal atomic functions (alternatively characterized by opposite pseudo-spin), and backscattering will be suppressed. We evaluate quantitatively the degree of the suppression in realistic III-V and II-VI quantum-well and wire structures and demonstrate that these are potentially useful in practical optoelectronic devices.

In the two-band approximation [16], the basis state of the heavy hole valence band (VB) can be written as $\left| V \right\rangle = 2^{-1/2} \left| X + jY \right\rangle f_{hh}(z)$, where X and Y are the bonding combinations of P-

type orbitals, and $f_{hh}(z)$ is the envelope function, while the conduction band (CB) state can be described as $|C\rangle = |jS\rangle f_c(z) + |jZ\rangle f_{lh}(z)$, where S is the anti-bonding combination of S-like orbitals, and Z is the light hole orbital, which is a bonding combination of P-type orbitals, $f_c(z), f_{lh}(z)$ are the envelope functions, normalized as $\int \left( f_c^2 + f_{lh}^2 \right) dz = 1$, and the matrix element of the in-plane momentum $\hat{\boldsymbol{p}}$ is $\boldsymbol{p}_{cv} = \langle C | \hat{\boldsymbol{p}} | V \rangle = 2^{-1/2} P_{cv} (\vec{x} + j\vec{y}) F_{cv}$, where $\vec{x}$ and $\vec{y}$ are the unity vectors, $P_{cv}$ is the Kane matrix element, and $F_{cv} = \int f_c(z) f_{hh}(z) dz$ is the overlap of envelope functions. The product of the above matrix element and the arbitrary in-pane wave vector becomes $\boldsymbol{k} \cdot \boldsymbol{p}_{cv} = 2^{-1/2} P_{cv} F_{cv} (k_x + j k_y) = 2^{-1/2} P_{cv} F_{cv} e^{j\theta_k}$, where $\theta_{\boldsymbol{k}}$ represents the wavevector direction.

By analogy with graphene, if we introduce the characteristic "Fermi" velocity $v_F = P_{cv} F_{cv} / \sqrt{2} m_0 \approx 10^8 F_{cv}$ cm/s, the Hamiltonian becomes:

$$H(\boldsymbol{k}) = \begin{pmatrix} E_g(k^2)/2 & \hbar k v_F e^{j\theta_k} \\ \hbar k v_F e^{-j\theta_k} & -E_g(k^2)/2 \end{pmatrix} = \hbar k_x v_F \sigma_x + \hbar k_y v_F \sigma_y + \frac{E_g(k^2)}{2} \sigma_z \qquad (1)$$

where $\boldsymbol{\sigma}$ is the Pauli matrix, and $E_g(k^2)$ is the term that includes the residual gap energy and the dispersive terms due to interactions with all the remote subbands, to be considered in more detail below. For $E_g = 0$, we obtain $H(\boldsymbol{k}) = \hbar v_F \boldsymbol{k} \cdot \boldsymbol{\sigma}$, and the solution is a perfect Dirac fermion with linear dispersion $E(k) = \pm \hbar v_F k$ and the wavefunction $|\boldsymbol{k}\rangle = 2^{-1/2} \left[ |C\rangle \mp |V\rangle e^{-j\theta_k} \right] e^{j\boldsymbol{k} \cdot \boldsymbol{r}}$, where the minus (plus) sign corresponds to the state with the higher (lower) energy corresponding to the CB (VB). One can now introduce a pseudo-spin with the eigenstates $\left|\Uparrow\right\rangle = \left| jS \right\rangle$ and $\left|\Downarrow\right\rangle = -\left| X + jY \right\rangle / \sqrt{2}$ and re-write the electronic state as:

$$\left| \boldsymbol{k} \right\rangle = \frac{1}{\sqrt{2}} e^{j\boldsymbol{k}\cdot\boldsymbol{r}} \begin{pmatrix} e^{j\theta_k/2} \\ be^{-j\theta_k/2} \end{pmatrix} \tag{2}$$

where $b = 1$ (-1) for the VB (CB). The dispersion relations in this approximation are exact replicas of those in graphene, albeit with a somewhat lower Fermi velocity if the overlap is significantly below unity. The overlap of the two states in the same band is easily calculated using $\left\langle \boldsymbol{k}_1 \middle| \boldsymbol{k}_2 \right\rangle = e^{j(\boldsymbol{k}_2 - \boldsymbol{k}_1)\cdot\boldsymbol{r}} \cos(\theta_{12}/2)$, where $\theta_{12} = \theta_{\boldsymbol{k}_1} - \theta_{\boldsymbol{k}_2}$ is the scattering angle. As expected, the overlap vanishes for two states with the opposite wave-vector, and only a short-range perturbation on the scale of a bond length can cause scattering, which precludes interface roughness and ionized impurity backward scattering and suppresses phonon scattering.

Unfortunately, the mobility is not dramatically enhanced because of the existence of other states. The momentum relaxation rate is proportional to $\tau_m^{-1}(k) \sim \int\limits_0^{2\pi} M^2(\boldsymbol{k}_1, \boldsymbol{k}_2)(1 - \cos\theta_{12}) d\theta_{12}$, where $M^2(\boldsymbol{k}_1, \boldsymbol{k}_2)$ is the scattering matrix element. For an ordinary III-V system, the matrix element depends only weakly on the direction of scattering, but for Dirac electrons it includes the overlap factor $\left| \left\langle \boldsymbol{k}_1 \middle| \boldsymbol{k}_2 \right\rangle \right|^2$. In the following, for definiteness we consider the case where the entire angular dependence of the scattering matrix element is given by the overlap factor. Hence, the ratio of the momentum scattering rate of Dirac electrons $\tau_{mD}^{-1}$ to that of ordinary electrons $\tau_{m0}^{-1}$ is given by:

$$\frac{\tau_{mD}^{-1}}{\tau_{m0}^{-1}} \sim \frac{\int\limits_{0}^{2\pi}(1-\cos\theta_{12})\cos^2\frac{1}{2}\theta_{12}d\theta_{12}}{\int\limits_{0}^{2\pi}(1-\cos\theta_{12})\ d\theta_{12}} = \frac{1}{4} \qquad (3)$$

Thus, as in the case of graphene, the electron mobility is enhanced only by a factor of four even in the simplified two-band formalism with complete suppression of backscattering and a flat angular dependence for the scattering matrix element apart from the overlap factor.

Let us now examine the band structure of the InAs/GaInSb/AlSb type-II QW [17,18] on a (001) GaSb substrate using the standard 8-band **k·p** calculations at $T = 10$ K [19]. In order to maximize the electron-hole wavefunction overlap while obtaining zero energy gap, the hole well is relatively thin (25 Å in the present case), while the thickness of the electron well is adjusted to give zero gap (71 Å). Furthermore, the wavefunction overlap is enhanced to be $F_{cv} = 12\%$ if the In content of the GaInSb layer is increased to the typical experimental maximum of 35% [20]. The zero-gap condition can be realized at higher $T$ by slightly decreasing the InAs thickness.

The dispersion relations for the electron (e) and hole (h) subbands near the Dirac point are shown in Fig. 1(b). The asymmetric nature of the InAs/GaInSb/AlSb QW leads to a pronounced Rashba spin splitting of the dispersions [21]. In spite of zero gap, the dispersion relations can be fitted linearly only in a very limited region of several tens of meV around the Dirac point, with the Fermi velocity of $v_F \approx 1.4\text{x}10^7$ cm/s, close to the expected value considering the reduced overlap. The e-h overlap can be improved by sandwiching a hole well between two 66-Å-thick electron wells in the so-called "W" structure [22]. $F_{cv}$ is estimated to be 25% or roughly double that for the InAs/GaInSb/AlSb QW, and the Fermi velocity increases to $v_F \approx$ 2.8x10$^7$ cm/s. The antisymmetric subband is separated from the Dirac point by 41 meV, which sets a temperature limit for this approach.

While III-V semiconductors are generally easier to grow and fabricate, we explore a type-I II-V QW system that is capable of reaching the gapless condition, namely, HgTe/HgCdTe, shown in Fig. 1(c). This occurs for a HgTe thickness of 63 Å and a thick $Hg_{0.15}Cd_{0.85}Te$ barrier [23, 24], with $F_{cv} \approx 68\%$, significantly larger than in the type-II structures. The electron dispersion in Fig. 1(d) is quite linear, which implies the absence of backscattering at higher temperatures, while the hole dispersions are complicated because of the coupling between the heavy- and light-hole subbands. Owing to a slightly lower accepted value of $P_{cv}$ in HgCdTe, $v_F \approx$ 6.0x10$^7$ cm/s, or ≈60% of that in graphene.

The overlap factor between the two states with opposite momentum $|<\mathbf{k}|-\mathbf{k}>|^2$ is plotted as a function of in-plane wavevector $\mathbf{k}$ in Fig. 2(a) for the three structures discussed above. For the InAs/GaInSb/AlSb QW, the overlap is small only near the Dirac point, while the "W" structure displays only limited improvement. At an in-plane wavevector of ≈0.06 x $2\pi/a$, the interactions with the lower-lying valence subbands produce a cancellation of the backscattering overlap for the symmetric "W" structure. On the other hand, the HgTe QW exhibits overlap factors <10$^{-3}$ for all the wavevectors of interest. The strength of backscattering is illustrated as a polar plot for two $\mathbf{k}$ values for the InAs/GaInSb/AlSb QW and the HgTe QW in Fig. 2(b) and (c), respectively. The former follows the theoretical $\cos^2(\theta/2)$ dependence only near the zone center.

The effect on the electron mobility expressed in terms of the momentum relaxation rate $\tau_m^{-1}$ normalized to the case of a unity overlap is shown in Fig. 3(a). While the magnitude of the reduction is a factor of 4 [as expected from Eq. (3)] near the Dirac point, it tends to decrease significantly for the asymmetric type-II structure and, to a lesser extent, for the "W" structure as **k** becomes larger owing to the interactions with other subbands. For large wavevectors, the CB becomes more energetically separated from the lower-lying subbands, and the normalized momentum scattering rate decreases once again. The HgTe QW allows a factor of 4-5 improvement to be realized at all **k** of interest. The normalized momentum scattering rate is also plotted vs. 2D electron density at $T = 10$ K in Fig. 3(b). For sheet carrier densities in the mid-$10^{11}$ cm$^{-2}$ range, the mobility may improve by a factor of $\approx$ 2, 3, and 6 for the InAs/GaInSb/AlSb, "W", and the HgTe QW structures, respectively. Furthermore, the mobility enhancement is expected to persist to liquid-helium and even ambient temperatures in the HgTe QW case.

In order to obtain a suppression of backscattering, it is not necessary for the energy gap to be precisely zero. In contrast to the case of graphene, the layer thicknesses must be adjusted to realize a vanishing gap when no field is applied along the growth direction. Nevertheless, even for energy gaps as wide as 30 meV, the suppression of backscattering is only $\approx$10% lower upon the angular averaging in Eq. (3).

While some issues arise in stacking multiple layers of graphene separated by the appropriate barriers, multiple QW of our Dirac structures can be grown epitaxially using well-known methods [25,26], with the corresponding scaling of the absorption strength. The material can act essentially as a bulk metal in the QW plane, while retaining its dielectric character in the growth direction. According to our preliminary estimates, the hyperbolic property [27] of this multi-QW material should allow volume plasma frequencies as high as 20 THz ($\lambda \approx 15$ μm).

While the electrons near the Dirac point of graphene have highly nonlinear characteristics, this is not the case at larger **k** [28]. Hence, one can achieve high, yet quickly saturating nonlinear susceptibility at low fields. For the structures considered here, this issue can be circumvented by utilizing numerous QWs to obtain large nonlinear phase shifts.

As shown above, the backscattering suppression of 2D Dirac electrons produces a significant, but limited increase in the mean free path and mobility. In order to achieve a more dramatic suppression, we consider a quasi-1D structure that allows the angular averaging to be eliminated. In contrast to Sakaki's proposal [7], backscattering is strictly forbidden for ideal Dirac electrons by the form of the wavefunction overlap determined above, and a large increase in the mobility is expected. The HgTe quantum "Dirac" wire with a cross section of 115 Å x 300 Å is modeled a 2D SL, with $Hg_{0.15}Cd_{0.85}Te$ barriers of 200 and 300 Å along the two respective orthogonal axes, as illustrated schematically in Fig. 4(a). The band structure for this SL with periodicity in 2D displays a nearly vanishing gap at the Dirac point, as shown in Fig. 4(b). The squared overlap $|<\boldsymbol{k}|\text{-}\boldsymbol{k}>|^2$ as a function of the electron wavevector $\boldsymbol{k}$ is plotted in Fig. 4(c). The initial value is due to the difficulty of adjusting the gap to be precisely zero. The carrier density per unit cell area of the SL and the corresponding average squared overlap in the low-temperature limit are shown in Fig. 4(d). The expected mobility enhancement by a factor of 30-50 is of the same order as in carbon nanotubes.

In conclusion, we have shown that zero-gap QWs and quantum wires made from III-V and II-VI semiconductors possess all the essential properties associated with Dirac electrons in graphene and carbon nanotubes, including pseudo-spin and suppressed scattering. Unlike graphene, typically limited to a single or few layers, the proposed structures can be epitaxially grown with numerous layers of Dirac electrons. Thus, all the benefits of Dirac electrons can be realized with not just high current densities, but also with high currents, and various possibilities for linear and nonlinear optical materials are enabled.

# Figure Captions

**Figure 1.** Band diagram, zone-center electron (dashed) and hole (dotted) wavefunctions, and band structure near the Dirac point for **(a), (b)** asymmetric type-II 71 Å InAs/25 Å $Ga_{0.65}In_{0.35}Sb$/AlSb QW on a (001) GaSb substrate, **(c), (d)** 63 Å HgTe/$Hg_{0.15}Cd_{0.85}Te$ type-I QW on a (001) CdZnTe substrate.

**Figure 2. (a)** The overlap factor $|<\mathbf{k}|\mathbf{-k}>|^2$ between two states with opposite momentum as a function of in-plane wavevector for the 63 Å HgTe/$Hg_{0.15}Cd_{0.85}Te$ QW (solid), 71 Å InAs/25 Å $Ga_{0.65}In_{0.35}Sb$/AlSb QW (dashed), 66 Å InAs/25 Å $Ga_{0.65}In_{0.35}Sb$/66 Å InAs/AlSb "W" QW structure (dotted); **(b)** Angular dependence of the scattering strength as a function of scattering angle for the 71 Å InAs/25 Å $Ga_{0.65}In_{0.35}Sb$/AlSb type-II QW, and **(c)** 63 Å HgTe/$Hg_{0.15}Cd_{0.85}Te$ type-I QW for the electron wavevectors of $10^{-2}$ and $10^{-3}$ x $2\pi/a$. The theoretical $\cos^2(\theta/2)$ dependence is indicated with a line for comparison.

**Figure 3.** Normalized momentum scattering rate as a function of **(a)** in-plane momentum and **(b)** 2D carrier density for the 71 Å InAs/25 Å $Ga_{0.65}In_{0.35}Sb$/AlSb QW (dashed), 66 Å InAs/25 Å $Ga_{0.65}In_{0.35}Sb$/66 Å InAs/AlSb "W" QW structure (dotted), and the 63 Å HgTe/$Hg_{0.15}Cd_{0.85}Te$ QW (solid).

**Figure 4**. **(a)** Cross-sectional schematic, **(b)** band structure, **(c)** overlap factor $|<\mathbf{k}|\mathbf{-k}>|^2$ between the two states with opposite momenta as a function of wavevector, **(d)** overlap factor vs. Fermi energy and volumetric carrier density for a nearly-zero-gap 115 Å x 300 Å HgTe/$Hg_{0.15}Cd_{0.85}Te$ multiple-quantum-wire array.

**References**

1. A. K. Geim, and K. S. Novoselov, Nature Mater. **6**, 183 (2007).

2. A. H. Castro-Neto, F. Guinea, N. M. R. Peres, K. S. Novoselov and A. K. Geim, Rev. Mod. Phys. **81**,109 (2009).

3. K. J-H Chen, C Jang, S. Xiao, M. Ishigami, and M. S. Fuhrer, *Nature Nanotechnology* **3**, 206 (2008).

4. M. Mecklenburg and B. C. Regan, Phys. Rev. Lett. **106**, 116803 (2011).

5. A. N. Calogeracos, and N. Dombey, Contemp. Phys. **40**, page number (1999).

6. V. W. Scarola, K. Park and S. Das Sarma, New J. Phys. **7**, 177 (2005).

7. H Sakaki, Jpn. J. Appl. Phys. **19**, L735 (1980).

8. C. L. Kane and E. J. Mele, Phys. Rev. Lett . **78**, 1932 (1997).

9. J-Y. Park, S. Rosenblatt, Y. Yaish, V. Sazonova, H. Ustunel, S. Braig, T. A. Arias, P. W. Brouwer, and P. L. McEuen, Nano Lett. **3**, 517 (2004).

10. D. Teweldebrhan, V. Goyal, and A. A. Balandin, Nano Lett. **10**, 1209 (2010).

11. D. Hsieh, D. Qian, L. Wray, Y. Xia, Y. S. Hor, R. J. Cava, and M. Z. Hasan, Nature **452**, 970 (2008).

12. C. Liu, T. L. Hughes, X-L. Qi, K. Wang, and S-C. Zhang, Phys. Rev. Lett **100,** 236601 (2008).

13. B. Büttner, C. X. Liu, G. Tkachov, E. G. Novik, C. Brüne, H. Buhmann, E. M. Hankiewicz,P. Recher, B. Trauzettel, S. C. Zhang and L.W. Molenkamp, Nature Physics, **7**, 418 (2011).

14. B. A. Bernevig, T. L. Hughes, and S-C. Zhang, Science **314**, 1757 (2006).

15. M. G. Silverinha, N. Engheta, Phys. Rev. B **86**, 161104(R) (2012)

16. G. Bastard, Phys. Rev. B **25**, 7584 (1982).

17. Y. Naveh and B. Laikhtman, Appl. Phys. Lett. **66**, 1980 (1995).

18. I. Knez and R.-R. Du, Front. Phys. **7**, 200 (2012).

19. R. Magri, L. W. Wang, A. Zunger, I. Vurgaftman, and J. R. Meyer, Phys. Rev. B **61**, 10235 (2000).

20. I. Vurgaftman, W. W. Bewley, C. L. Canedy, C. S. Kim, M. Kim, J. R. Lindle, C. D. Merritt, J. Abell, and J. R. Meyer, IEEE J. Select. Top. Quantum Electron. **17**, 1435 (2011).

21. Y. A. Bychkov and E. I. Rashba, J. Phys. C **17**, 6039 (1984).

22. J. R.Meyer, C. A. Hoffman, F. J. Bartoli, and L. R. Ram-Mohan, Appl. Phys. Lett. **67**, 757 (1995).

23. M. König, S. Wiedmann, C. Brüne, A. Roth, H. Buhmann, L. W. Molenkamp, X.-L. Qi, and S.-C. Zhang, Science **318**, 767 (2007).

24. M. König, H. Buhmann, L. W. Molenkamp, T. Hughes, C.-X. Liu, X.-L. Qi, and S.-C. Zhang, J. Phys. Soc. Japan **77**, 031007 (2008).

25. C. L. Canedy, J. Abell, W. W. Bewley, E. H. Aifer, C. S. Kim, J. A. Nolde, M. Kim, J. G. Tischler, J. R. Lindle, E. M. Jackson, I. Vurgaftman, and J. R. Meyer, J. Vac. Sci. Technol. B **28**, C3G8 (2010).

26. S. Dvoretsky, N. Mikhailov, Y. Sidorov, V. Shvets, S. Danilov, B. Wittman, and S. Ganichev, J. Electron. Mater. **39**, 918 (2010).

27. V. Drachev, V. A. Podolskiy, and A. V. Kildishev, Opt. Express **21**, 15048 (2013).

28. S. A. Mikhailov and K. Ziegler, J. Phys.: Condens. Matter **20**, 384204 (2008).

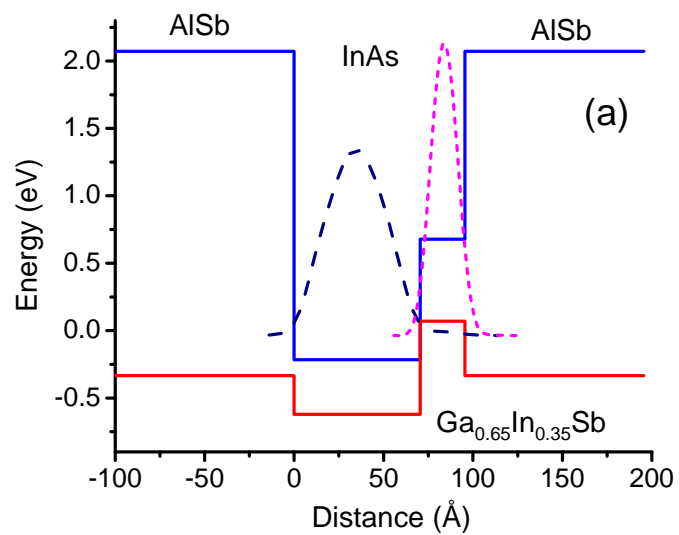

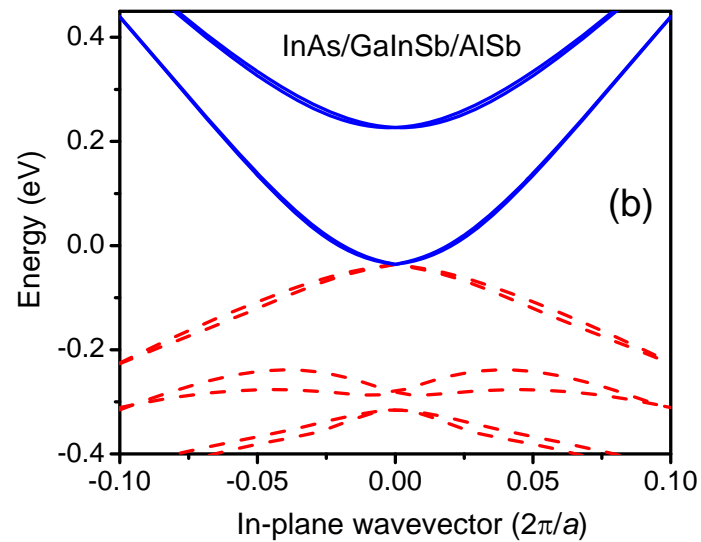

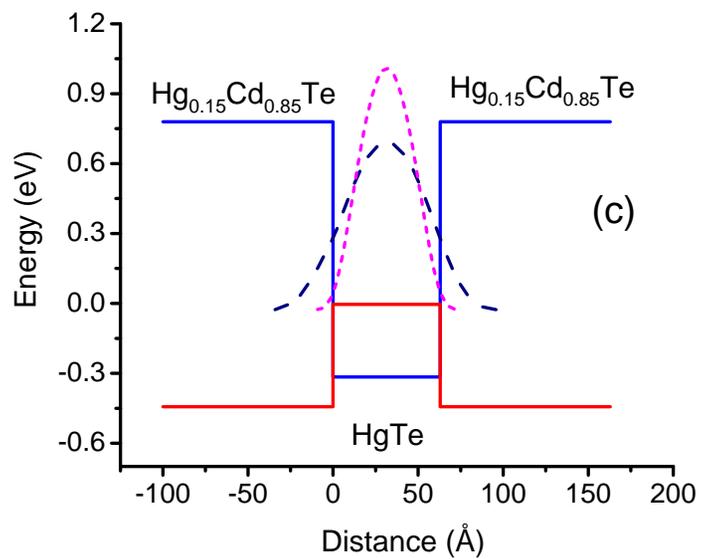

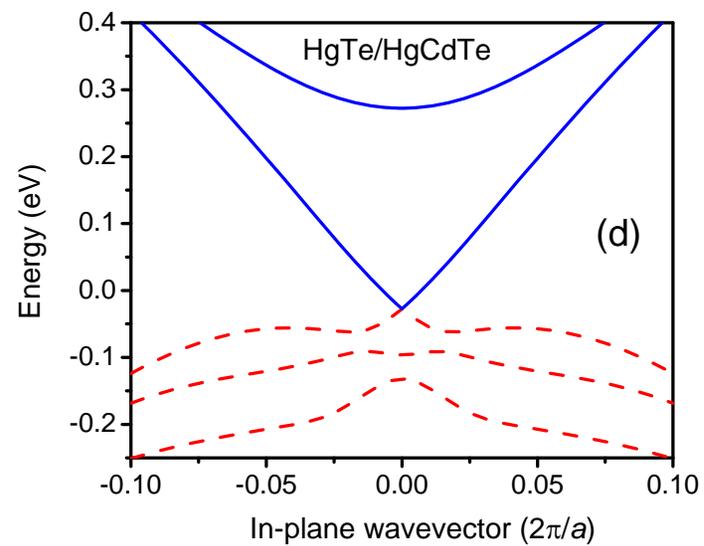

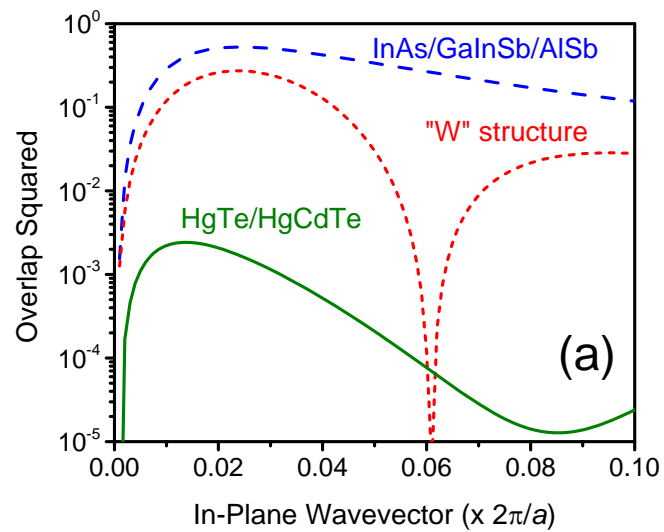

InAs/GaInSb/AlSb

"W" structure

HgTe/HgCdTe

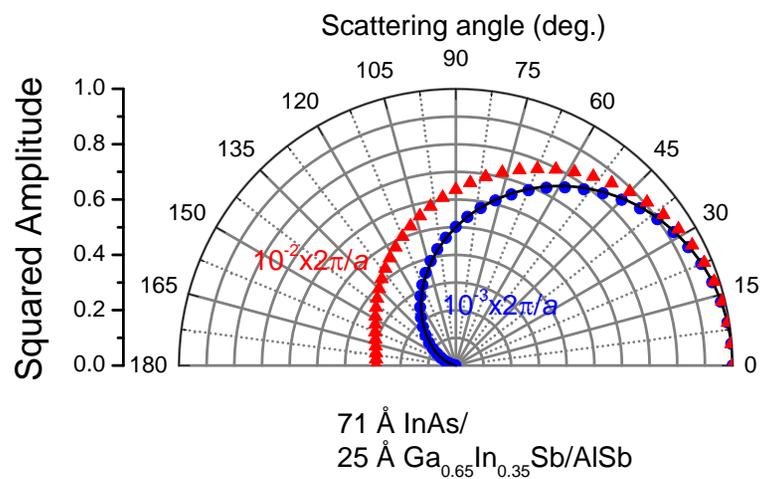

71 Å InAs/
25 Å Ga$_{0.65}$In$_{0.35}$Sb/AlSb

(b)

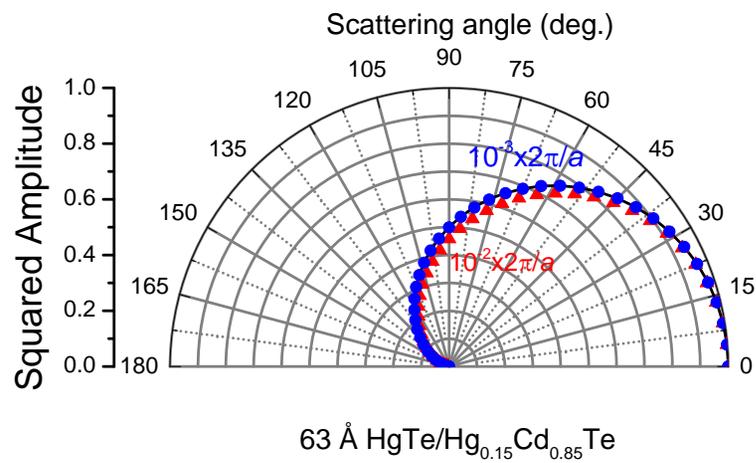

63 Å HgTe/Hg$_{0.15}$Cd$_{0.85}$Te

(c)

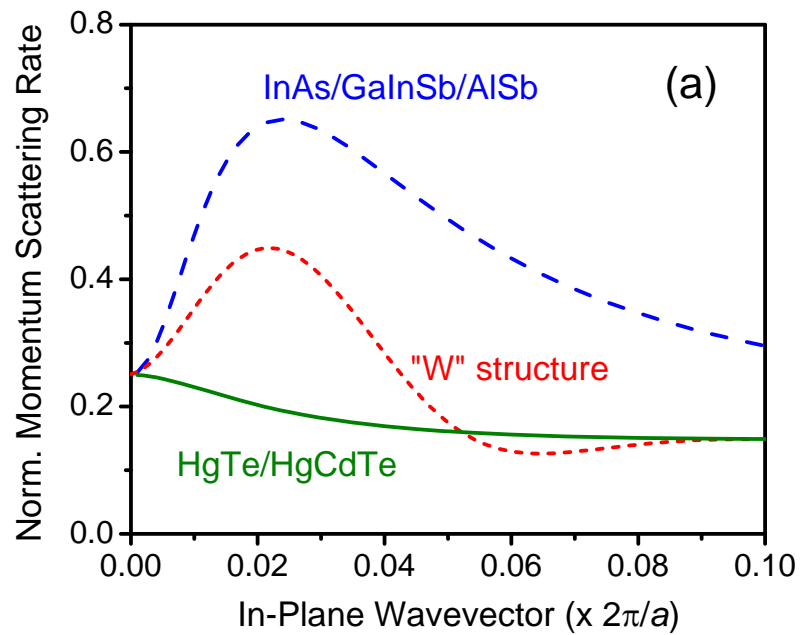
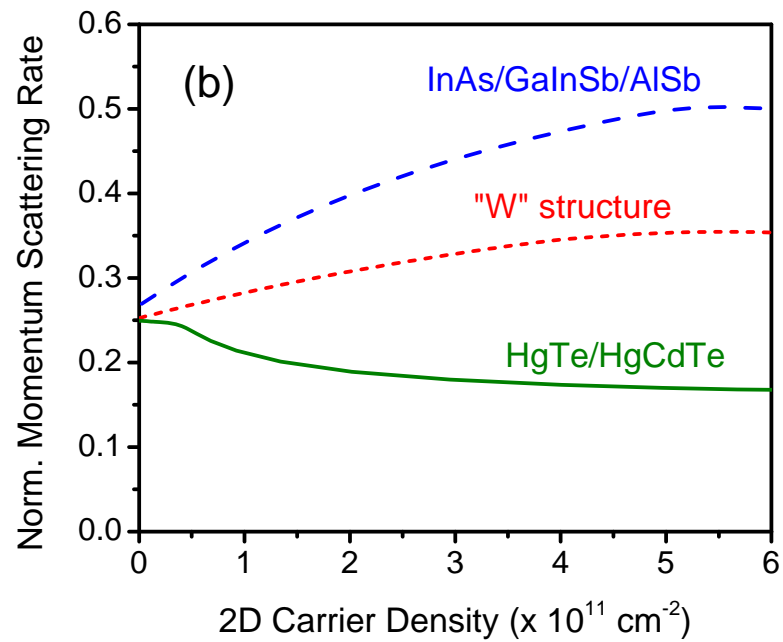

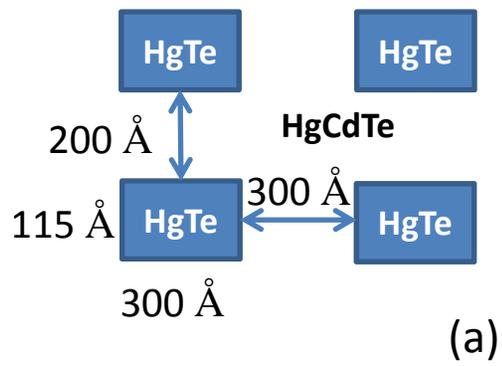

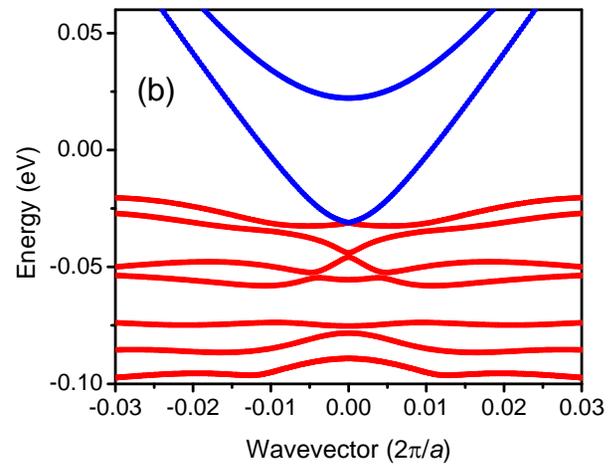

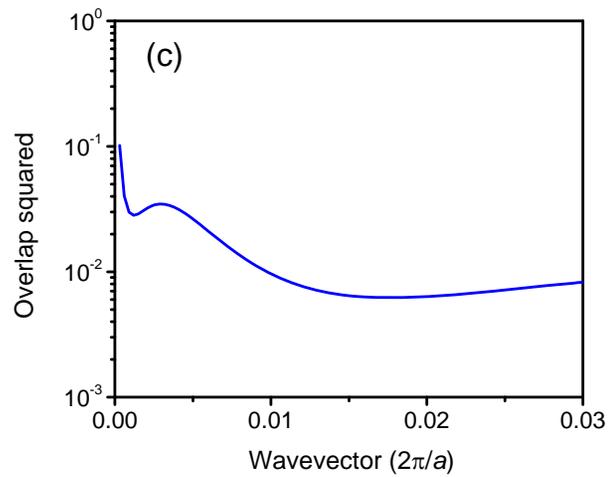

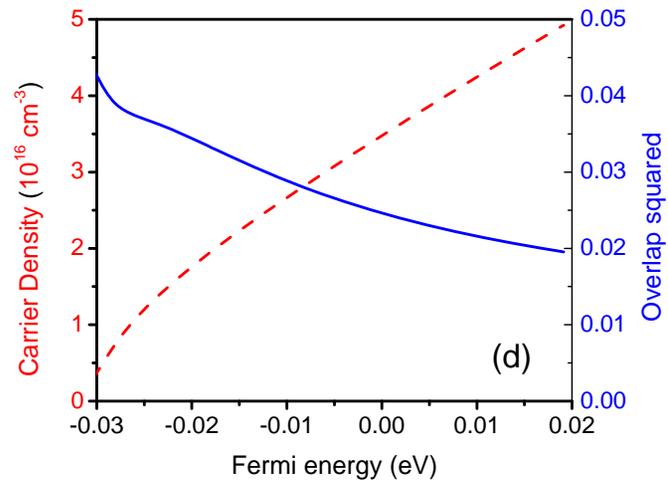